\pdfoutput=1
\documentclass[a4paper]{jpconf}
\usepackage{graphicx,subfigure}
\newcommand{\be}{\begin{eqnarray}}
\newcommand{\ee}{\end{eqnarray}}

\renewcommand{\sc}{\slashchar}

\def\slashchar#1{\setbox0=\hbox{$#1$}           
   \dimen0=\wd0                                 
   \setbox1=\hbox{/} \dimen1=\wd1               
   \ifdim\dimen0>\dimen1                        
      \rlap{\hbox to \dimen0{\hfil/\hfil}}      
      #1                                        
   \else                                        
      \rlap{\hbox to \dimen1{\hfil$#1$\hfil}}   
      /                                         
   \fi}                                         %

\begin{document}
\title{Monte-Carlo simulations of QCD Thermodynamics in the PNJL model}

\author{M Cristoforetti, T Hell and W Weise}

\address{Physik-Department T39, Technische Universit\"at M\"unchen, James-Franck-Str. 1, D-85747 Garching}
\address{}
\address{\textit{Work supported in part by BMBF, GSI and the DFG Excellence Cluster ``Origin and Structure of the Universe'' }}

\ead{mcristof@ph.tum.de}

\begin{abstract}
We apply a Monte-Carlo method  to the two flavor Polyakov loop extended Nambu and Jona-Lasinio (PNJL) model. In such a way we can go beyond mean field calculations introducing fluctuations of the fields. We study the impact of fluctuations on the thermodynamics of the model. We calculate the second derivatives of the thermodynamic grand canonical partition function with respect to the chemical potential and present a comparison with lattice data also for flavor non-diagonal susceptibilities.
\end{abstract}

\section{Introduction}
Results of QCD thermodynamics from lattice computations are reproduced surprisingly well with a quasiparticle model, an extension of the Nambu -- Jona-Lasinio model with inclusion of Polyakov loop dynamics (the PNJL model) at the mean field level~\cite{Fukushima:2003fm,Fukushima:2003fw,Hatta:2003ga,Ratti:2005jh,Roessner:2006xn,Ratti:2007jf,Rossner:2007ik,Hell:2008cc}. A better understanding of these results requests the investigation of fluctuations in the PNJL model. This can be done by numerical simulations of the thermodynamics using standard Monte-Carlo techniques (MC-PNJL). The advantage over other methods is that Monte-Carlo calculations automatically incorporate fluctuations to all orders. 

We present new results of such computations and discuss the role and importance of fluctuations beyond mean field. It turns out that the temperature dependence of the second non-diagonal derivative of the thermodynamic grand canonical partition function with respect to quark chemical potentials is particularly sensitive to such fluctuations. While the mean field PNJL  predicts a vanishing coefficient, our calculation agrees well with available lattice data.

\section{The PNJL partition function}\label{sec:PNJLpf}
The Euclidean action of the two-flavor PNJL model, at finite baryon and isospin chemical potential, is given by \cite{Zhang:2006gu,Rossner:2007ik}
\be\label{eq:act}
	\nonumber\mathcal{S}_{E}(\psi,\bar{\psi},\phi)&=&\int_0^{\beta=1/T}\textrm{d}\tau\int\textrm{d}^3x\ [\bar{\psi}(i\sc{D}+\gamma_0\tilde{\mu}-m_0)\psi\\	
	&&+G((\bar{\psi}\psi)^2+(\bar{\psi}i\gamma_5\vec{\tau}\psi)^2]-\beta\int\textrm{d}^3x\ \mathcal{U}(\phi,\beta),
\ee
where $\psi$ is the $N_f=2$ doublet quark field and $m_0=\textrm{diag}(m_u,m_d)$ is the quark mass matrix.
The quark chemical potential matrix is defined as $\tilde{\mu}=\textrm{diag}(\mu+\mu_I,\mu-\mu_I)$, where $\mu_I$ is the isospin isovector chemical potential. 
In the PNJL model quarks interact with a background color gauge field $A_4=iA_0$, where $A_0=\delta_{\mu 0}g\mathcal{A}_a^{\mu}t^a$ with $\mathcal{A}_a^{\mu}\in \textrm{SU}(3)_c$ and $t^a=\lambda^a/2$. The field $A_4$ is related to the traced Polyakov loop by:
\be\label{eq:pl}
	\Phi=\frac{1}{N_c}\textrm{tr}_c L &\textrm{with}\ L=\exp\left(i\int_0^{\beta}\textrm{d}\tau A_4\right)& \textrm{and}\ \beta=\frac{1}{T}.
\ee
In the Polyakov gauge, the matrix $L$ is given in a diagonal representation
\be 
	L=\exp(i(\phi_3\lambda_3+\phi_8\lambda_8)).
\ee
The dimensionless effective fields $\phi_3$ and $\phi_8$ introduced here are identified with the Euclidean gauge fields in temporal direction divided by the temperature, $A_4^{(3)}/T$ and $A_4^{(8)}/T$. These two fields are a parametrisation of the diagonal elements of $\textrm{SU}(3)_c$.

The effective potential $\mathcal{U}$ models the confinement-deconfinement transition in pure gauge QCD  on mean-field Ginzburg-Landau level. 
In this paper we consider an ansatz for the effective potential given in \cite{Roessner:2006xn,Ratti:2006wg} motivated by the $\textrm{SU}(3)$ Haar measure:
\be\label{eq:effpphi}
	\frac{\mathcal{U}(\Phi,\Phi^*,T)}{T^4}=-\frac{1}{2}a(T)\Phi^*\Phi+b(T)\ln[1-6\Phi^*\Phi+4(\Phi^{*^3}+\Phi^3)-3(\Phi^*\Phi)^2],
\ee
where the temperature dependent prefactors are given by
\be
	a(T)=a_0+a_1\left(\frac{T_0}{T}\right)+a_2\left(\frac{T_0}{T}\right)^2 &\textrm{and}& b(T)=b_3\left(\frac{T_0}{T}\right)^3.
\ee
The parameters are chosen such that the critical temperature of the first order transition is equal to $T_0=270$ MeV, as given by lattice calculations, and that $\Phi^*,\Phi\rightarrow 1$ as $T\rightarrow\infty$.

After performing a bosonization of the PNJL Lagrangian introducing the auxiliary fields $\sigma$ and $\pi$ the partition function can be rewritten as
\be
	\mathcal{Z}&=&\mathcal{N}\int\mathcal{D}\phi\mathcal{D}\sigma\mathcal{D}\pi\exp[-\mathcal{S}[\sigma(\vec{x}),\pi(\vec{x}),\phi(\vec{x})]]\nonumber\\
	&=&\mathcal{N}\int\mathcal{D}\phi\mathcal{D}\sigma\mathcal{D}\pi\exp\Big[\frac{1}{2}\textrm{Tr}\ln[S^{-1}]-\beta\int\textrm{d}^3x\Big(\mathcal{U}(\phi,\beta)+\Big(\frac{\sigma^2+\vec{\pi}^2}{2G}\Big)\Big)\Big].
\ee
with the inverse quark propagator 
\begin{displaymath}
	S^{-1}=
	\left(\begin{array}{cc}
		-\sc{\partial}+(\mu+\mu_I-iA_4)\gamma_0-M& i\gamma_5 \pi\\
		i\gamma_5 \pi& -\sc{\partial}+(\mu-\mu_I-iA_4)\gamma_0-M
		\end{array}
	\right),
\end{displaymath}
and	$M=m_0-\sigma$.

\begin{figure}[ht!]
        \centering
        \includegraphics[width=.6\textwidth]{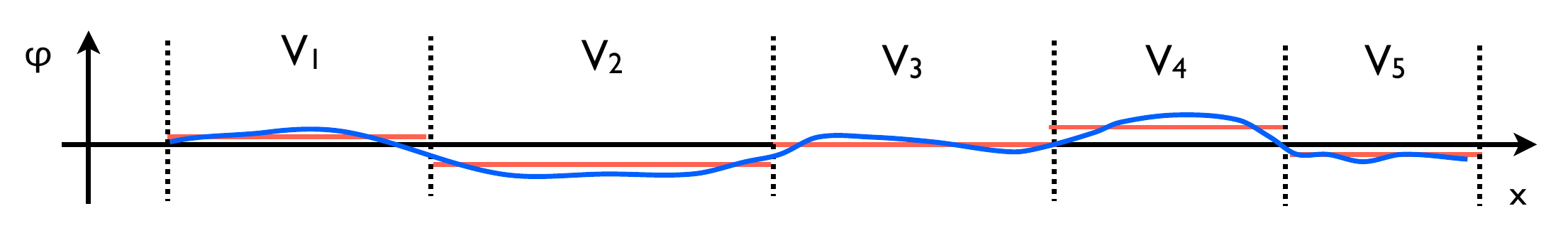}
      \caption{Approximation of the space dependence of the fields: the behavior of the field (blue line) is approximated in our model using a mean value (red lines) on subvolumes. In this way we permit small fluctuations of the field.}\label{fig:approx}
\end{figure}

In order to introduce fluctuations in the PNJL model we consider the total volume of our system as composed of a certain number of subregions, $V=\sum_i V_i$, and require that the fields are completely correlated if they lie in the same subvolume, and decorrelated otherwise \cite{Megias:2006bn}. In this way the fields can fluctuate in the sense that they can have different values from one subvolume to another. Physically this means that our fields are only weakly dependent on the position and can be considered constant inside subvolumes of a given dimension (Fig.~\ref{fig:approx}). 

From this assumption it follows that the action $\mathcal{S}[\sigma(\vec{x}),\pi(\vec{x}),\phi(\vec{x})]$ can be decomposed into a sum of contributions related to the $i$-th volume 

\be
	\mathcal{S}[\sigma(\vec{x}),\pi(\vec{x}),\phi(\vec{x})]=\sum_i\left.\mathcal{S}[\sigma(\vec{x}),\pi(\vec{x}),\phi(\vec{x})]\right|_{\vec{x}\in V_i}=\sum_i\mathcal{S}_i[\sigma,\pi,\phi]. 
\ee

The last equality emphasizes the strong correlation of the fields inside each subvolume. We assume therefore that the fields are constant for all $\vec{x}\in V_i$, hence, the partition function becomes 
\be\label{eq:pfuncvi}
	\mathcal{Z}&=&\prod_i\Big(\int\mathcal{D}\phi_i\mathcal{D}\sigma_i\mathcal{D}\pi_i\exp[-\mathcal{S}_i(\sigma,\pi,\phi)]\Big).
\ee

Computing the partition function~(\ref{eq:pfuncvi}) we introduce fluctuations in the PNJL model. 

\section{Monte-Carlo method}\label{sec:MC}
In order to evaluate (\ref{eq:pfuncvi}), we adopt standard Monte-Carlo techniques. Using the Metropolis algorithm we start from a random configuration $C$ and suggest a new configuration $C'$ that will be accepted with probability
\be
	p=\textrm{min}\{1,\exp(-S[C'])/\exp(-S[C])\}.
\ee

Writing Eq.~(\ref{eq:pfuncvi}), we point out that the fields involved in each subvolume are independent from the fields in all the other volumes. As a consequence, if we update a given configuration moving only the fields in the $k$-th subregion the action changes according to 
\be
	\nonumber  S(\phi_1,...,\phi'_k,...\phi_n)&=&\sum_{i\neq k}^n S(\phi_i)+S(\phi'_k) \\
	\nonumber S(\phi_1,...,\phi_k,...,\phi_n)&=&\sum_{i=1}^n S(\phi_i) \\
	S(\phi_1,...,\phi'_k,...,\phi_n)-S(\phi_1,...,\phi_k,...,\phi_n)&=&S(\phi'_k)-S(\phi_k)
\ee
this means that the Metropolis algorithm acts independently on each single volume. If we thus assume that every $V_i$ will be of a fixed size, $V_i=\tilde{V}$, Eq.~(\ref{eq:pfuncvi}) can be rewritten in momentum space as
\be\label{eq:pf}
	\mathcal{Z}&=&\Big(\int\mathcal{D}\phi\mathcal{D}\sigma\mathcal{D}\pi\exp\Big[\beta\tilde{V}\frac{1}{2}\sum_n\int\frac{\textrm{d}^3p}{(2\pi)^3}\textrm{Tr}\ln[ S^{-1}(i\omega_n,\vec{p})]\nonumber\\
	&&-\beta \tilde{V}\Big(\mathcal{U}(\phi,\beta)+\Big(\frac{\sigma^2+\pi^2}{2G}\Big)\Big)\Big]\Big)^n\hspace{1cm} n\rightarrow\infty.
\ee
The calculation of the partition function using Monte-Carlo can be performed by generating a set of configurations for a given subvolume. We only need to fix the size of this volume. Here we adopt the standard temperature dependence of the volume used in lattice calculations,
\be
	a=\frac{1}{N_t T}& \rightarrow &V=N_s^3 a^3=\frac{N_s^3}{N_t^3 T^3},
\ee
where $a$ is the lattice spacing, $N_t$ is the number of lattice sites in the Euclidean time direction and $N_s$ the number of lattice sites in the space dimensions.
It follows that 
\be\label{eq:vk}
	V\propto k/T^3,
\ee
where $k$ will be conveniently chosen. In this paper we perform calculations at different values of $k$ in order to understand the impact of $k$ on thermodynamic quantities.
In the next sections we will see how this Monte-Carlo version of the PNJL model (MC-PNJL) works both in pure gauge and in the $N_f=2$ case.

\section{Pure Gauge MC-PNJL}\label{sec:PG}
In this section we apply the Monte-Carlo approach to pure gluodynamics, i.e. to the Polyakov loop sector of the model. Contributions due to fluctuations in this case are very small, although we prefer to start with this simple case, because it is instructive for the understanding of the method.

\begin{figure}[ht!]
        \centering
        \subfigure{\includegraphics[width=.45\textwidth]{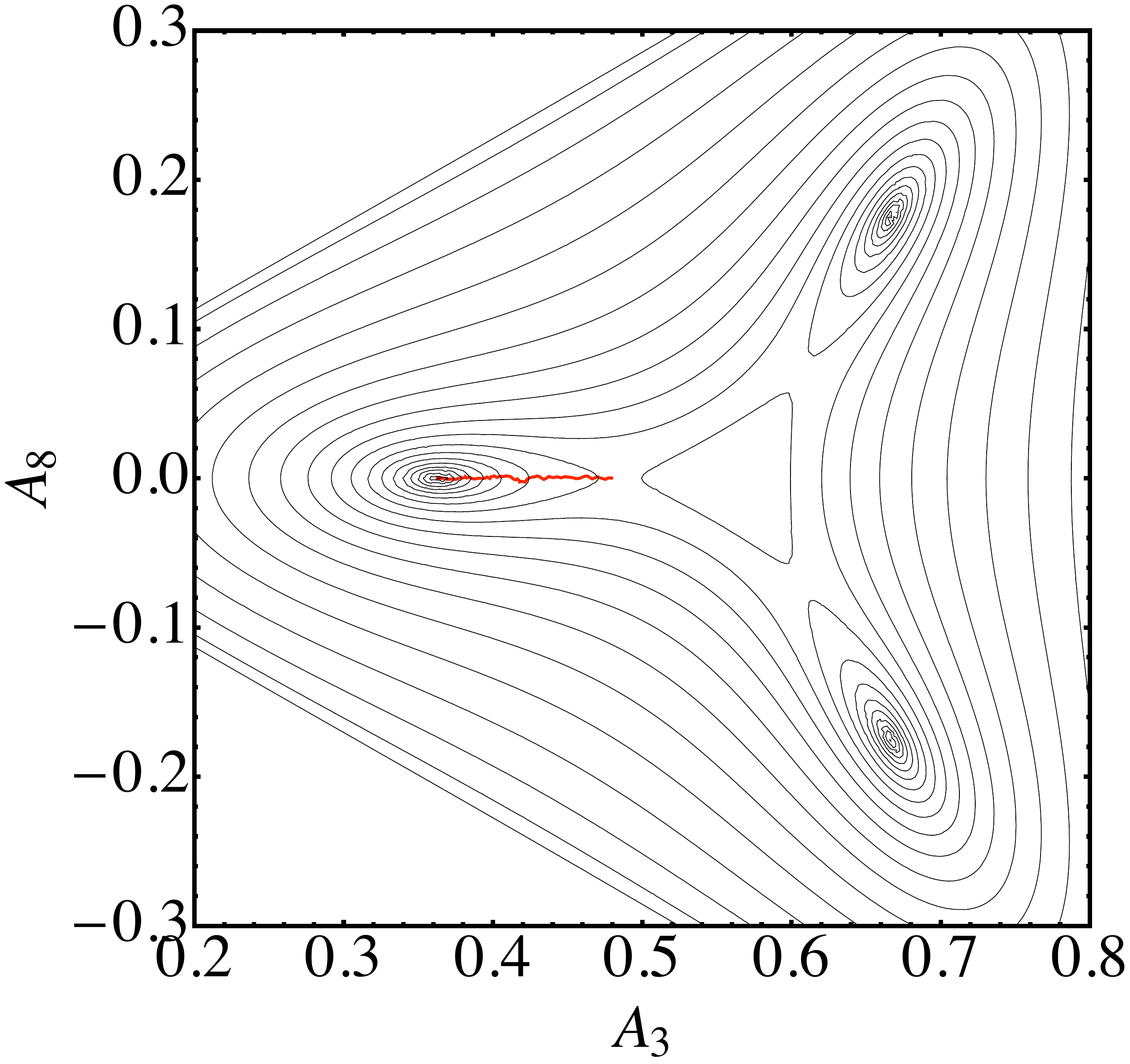}}\hspace{5mm}%
        \subfigure{\includegraphics[width=.45\textwidth]{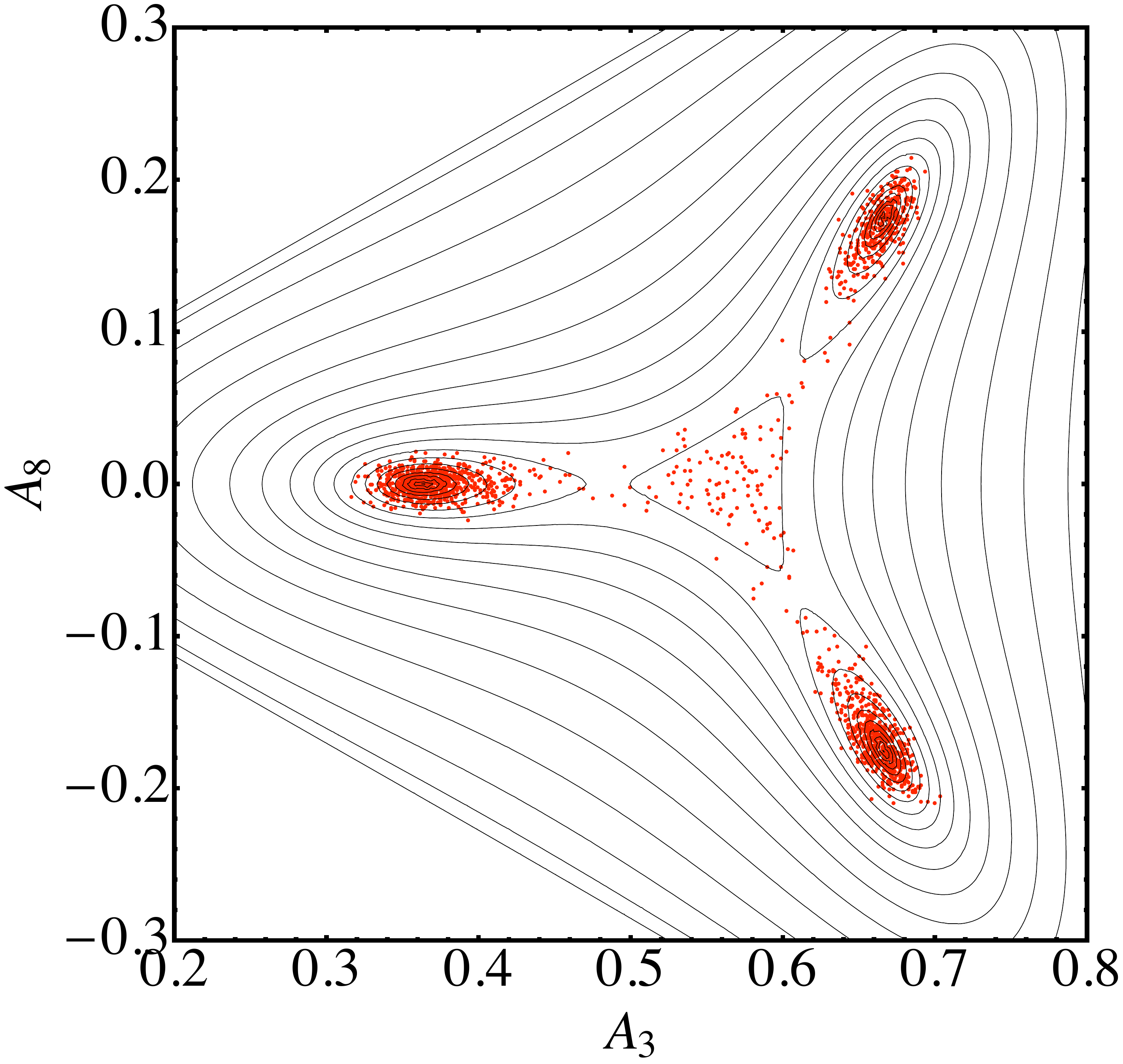}}
      \caption{Comparison between the cooling (left) and Metropolis (right) procedure applied to the Polyakov loop effective potential at $T=T_c=0.27$ GeV.}\label{fig:CoolvsMetr}
\end{figure}

We start with a generic configuration that in our system corresponds to $A_3$ and $A_8$ at a given temperature. Using Monte-Carlo we generate points in the configuration space close to the minimum of the action. In particular applying cooling we will reach exactly the minimum. In Fig.~\ref{fig:CoolvsMetr} we can see the difference between cooling and Metropolis in the case of $T=T_c=0.27$ GeV. For cooling (left picture) the action must decrease, and in few steps we fall into one of the three minima of the potential. On the other hand, using the full Metropolis the action can also slightly increase. As a consequence, the space of the available configurations is much larger (right picture). 

\begin{figure}[ht!]
      \centering
       \subfigure{\includegraphics[width=.44\textwidth]{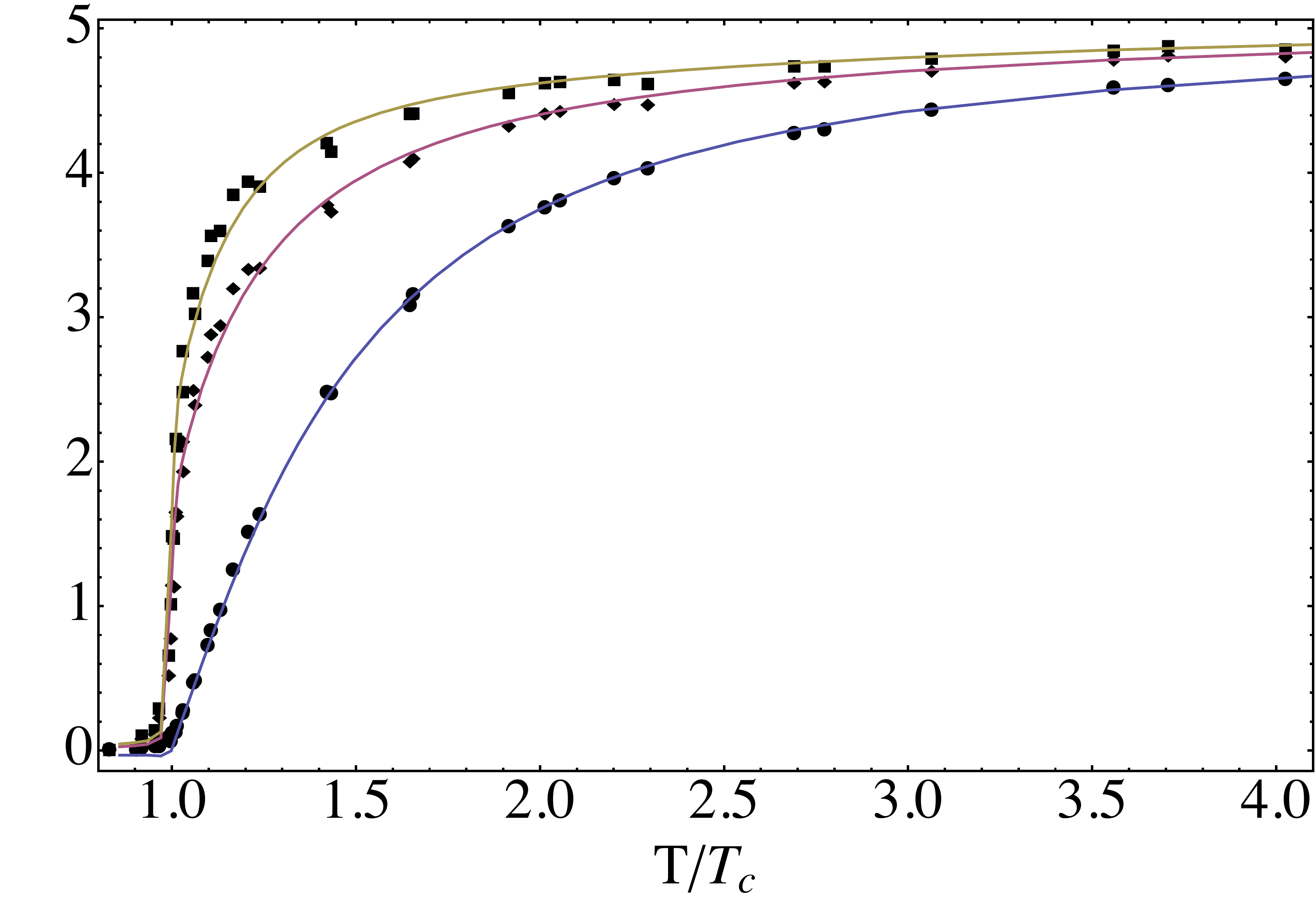}}\hspace{5mm}%
        \subfigure{\includegraphics[width=.45\textwidth]{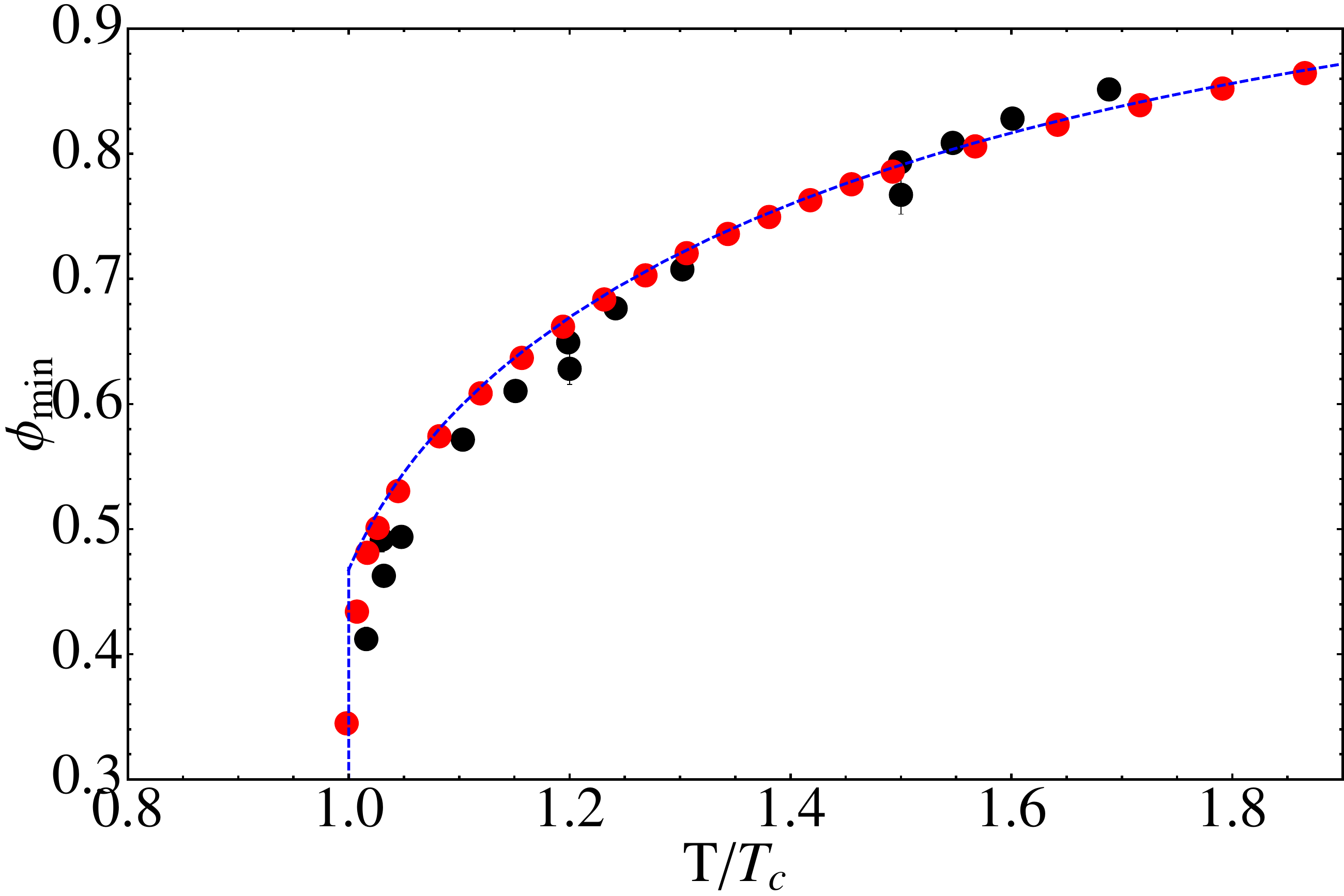}}
      \caption{Left: fit of Lattice data \cite{Boyd:1996bx} for pressure, energy and entropy density. Right: Comparison of the Polyakov loop dynamics with the data in  \cite{Kaczmarek:2002mc}}\label{fig:peePG}
\end{figure}

To fix the parameters of the effective potential we have computed the pressure, energy and entropy density at different temperatures including fluctuations, trying to fit the available lattice data (Fig.~\ref{fig:peePG} left). In the right panel of Fig.~\ref{fig:peePG} we compare the Polyakov loop obtained using the mean-field method (dashed line) with our results (red points) and the lattice data (black points) we can see that there is practically no difference between the two methods: both agree well with lattice data.

Once established how this method works in the case of the pure gluodynamics we can move to the more interesting case of the two flavors MC-PNJL model.

\section{Two flavors MC-PNJL model}\label{sec:2nf}
The starting point studying the thermodynamics for $N_f=2$ is the partition function (\ref{eq:pf}). The degrees of freedom in this case are the $A_3$ and $A_8$ components of the gauge field, and the two bosonic fields $\sigma$ and $\pi$. As explained before, evaluating the path integral, we need to fix the size of the volume $\tilde{V}$ as a function of the temperature. According to Eq.~(\ref{eq:vk}) we only have to fix the value of $k$. Since the value of this parameter is not known a priori, in this paper we consider four different values, $k=95,135,500,5000$. In this way we can study the dependence of the observables as a function of the volume size.

\subsection{Chiral and deconfinement transitions}
\begin{figure}[ht!]
        \centering
        \includegraphics[width=.6\textwidth]{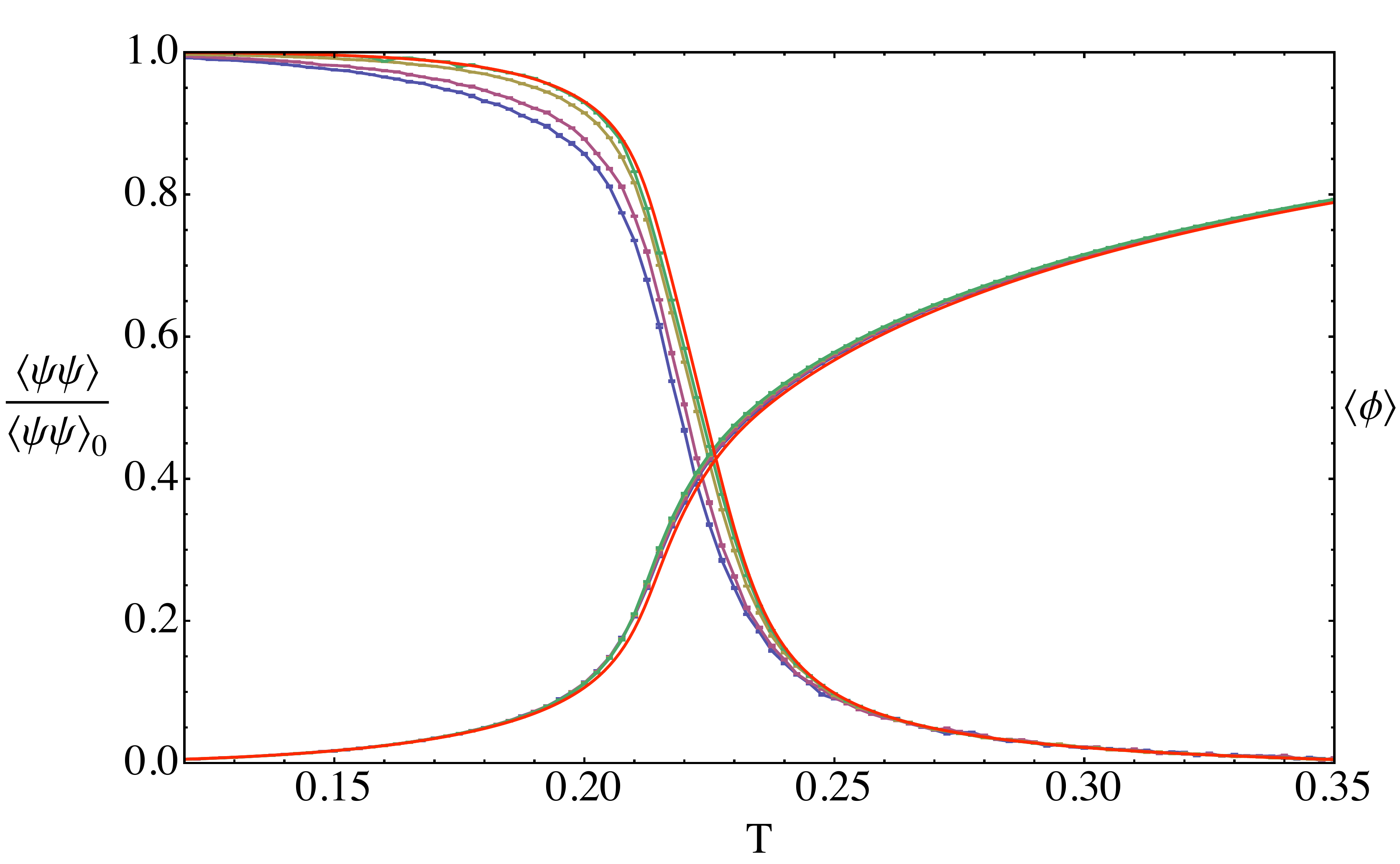}
      \caption{Chiral condensate and Polyakov loop: dependence on the volume size. The red lines are the mean field calculation, while moving from the blue to the green line we go from $k=95$ to $k=5000$.}\label{fig:comop}
\end{figure}

Since our aim is the investigation of the QCD phase diagram, the first thing we need to study is how the chiral and deconfinement transitions are affected by the introduction of fluctuations in the model. This can be achieved evaluating the chiral condensate and the Polyakov loop for the different volumes and comparing that with the mean-field result. This comparison is presented in Fig.~\ref{fig:comop}:  the presence of fluctuations does not modify the behavior of the Polyakov loop, the five different sets of data overlap perfectly. For the chiral condensate we can notice only small dependence on the temperature below the critical temperature. In analogy to the pure gauge case we can conclude for $N_f=2$ that the order parameters of the transitions are not affected by the presence of fluctuation. 

This last observation is indeed important when we move to the case of finite chemical potential, because this means that we can use mean-field approximation to determine the critical temperature: corrections due to the presence of fluctuations are negligible, and this permits to overcome the sign problem which is not present in mean field calculations.

\section{Taylor expansion at finite chemical potential}\label{sec:pressure}
The Taylor expansion approach to QCD thermodynamics at non-zero quark chemical potential represents one of the possibilities to overcome the sign problem on the lattice: instead of performing explicit calculations at $\mu_q\neq0$ one expands the thermodynamic potential in a Taylor series around zero chemical potential,
\be
	\frac{\Omega(T,\mu)}{T^4}=\frac{1}{VT^3}\ln\mathcal{Z}=\sum_{n=0}^{\infty}c_n(T)\Big(\frac{\mu}{T}\Big)^n,
\ee
with
\be
	c_n(T)=\frac{1}{n!}\frac{\partial^n\Omega}{\partial(\mu/T)^n}\Big|_{\mu=0}=\frac{1}{n!VT^3}\frac{\partial^n\ln\mathcal{Z}}{\partial(\mu/T)^n}\Big|_{\mu=0}
\ee
and $n$ even due to the $CP$ symmetry of the system.

The comparison between lattice data and MC-PNJL calculations for these coefficients represents a crucial test of the model. In particular the second non-diagonal coefficient is the most important because, as explained in the introduction, to obtain a non-vanishing result for $c_2^{ud}$ we must take into account fluctuations of the fields. We know that the size of fluctuations in our model depends on the size of our correlation volume, therefore, as before, we evaluate the Taylor coefficients for different values of the parameter $k$.

To perform the calculations using Monte-Carlo, we need to represent the derivative of the logarithm of the partition function in an explicit form. To do that, we start from the definition of our partition function,
\be
	\mathcal{Z}[T,\mu,\mu_I]=\int\mathcal{D}f\mathcal{D}A\exp\Big[\frac{V}{T}\ln\det M[T,\mu,\mu_I,f,A]-S_g[A]\Big].
\ee
From this expression we have
\be
	\frac{\partial\ln\mathcal{Z}[T,\mu,\mu_I]}{\partial\mu}&=&\frac{\partial}{\partial\mu}\ln\int\mathcal{D}f\mathcal{D}A\exp\Big[\frac{V}{T}\ln\det M[T,\mu,\mu_I,f,A]-S_g[A]\Big]\nonumber\\
	&=&\frac{1}{\mathcal{Z}[T,\mu,\mu_I]}\frac{V}{T}\int\mathcal{D}f\mathcal{D}A\frac{\partial\ln\det M[T,\mu,\mu_I,f,A]}{\partial\mu}e^{-S[T,\mu,\mu_I,f,A]}\nonumber\\
	&=&\frac{V}{T}\Big\langle\frac{\partial\ln\det M[T,\mu,\mu_I,f,A]}{\partial\mu}\Big\rangle.
\ee
Proceeding in the same way for the second derivative we obtain for the $c_2$ and isovector $c_2^I$ coefficients
\be
	c^{(I)}_2(T)&=&\frac{1}{2}\frac{T^2}{VT^3}\frac{\partial^2\ln\mathcal{Z}[T,\mu,\mu_I,f,A]}{\partial\mu_{(I)}^2}\nonumber\\
	&=&\frac{1}{2T^2}\Big\langle\frac{\partial^2\ln\det M[T,\mu,\mu_I,f,A]}{\partial\mu_{(I)}^2}\Big\rangle,
\ee
where we have neglected the expectation values which involves odd derivatives, due to the symmetry property of the determinant.  

\begin{figure}[ht!]
       \centering
        \subfigure{\includegraphics[width=.44\textwidth]{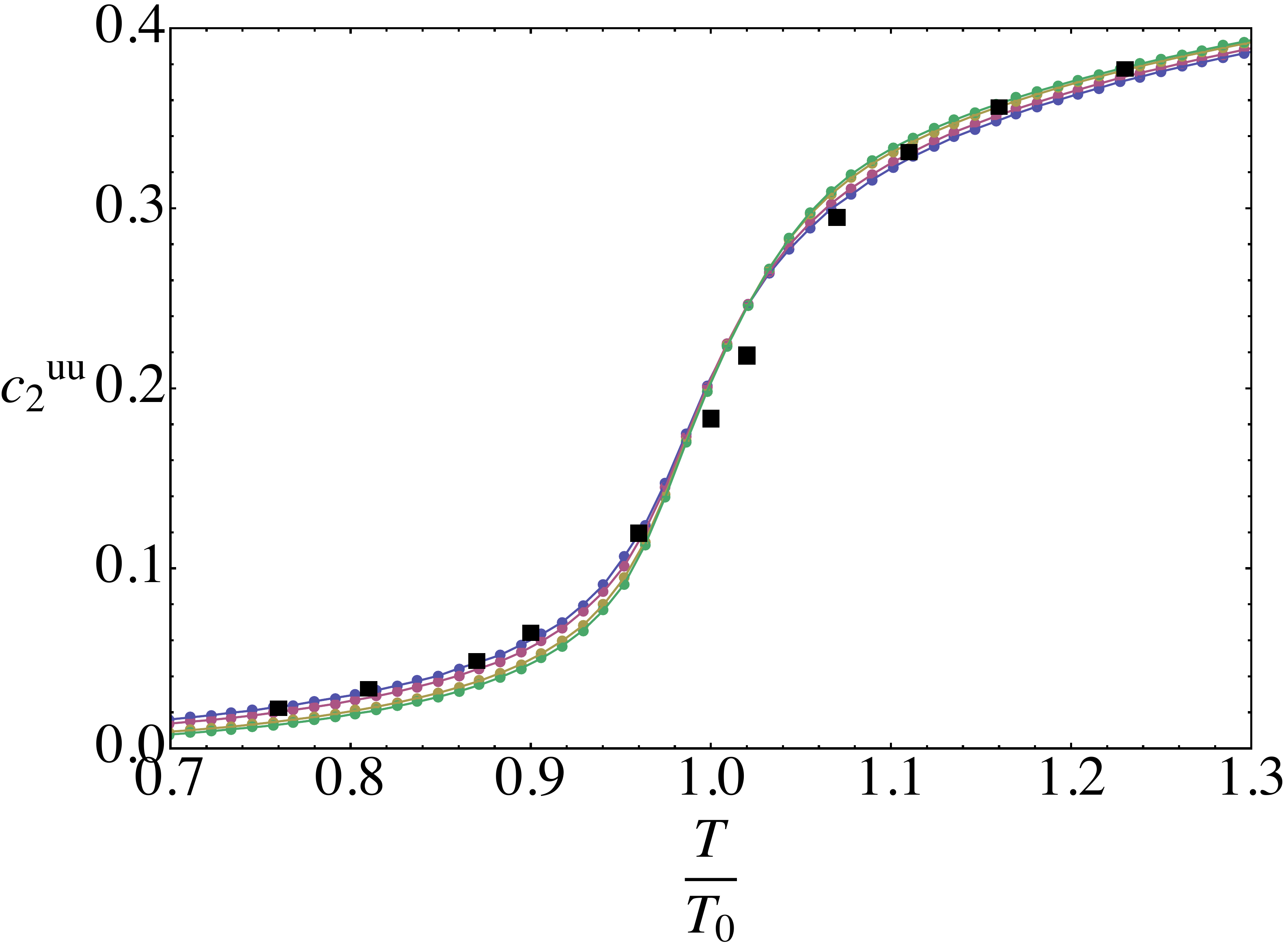}}\hspace{5mm}%
        \subfigure{\includegraphics[width=.5\textwidth]{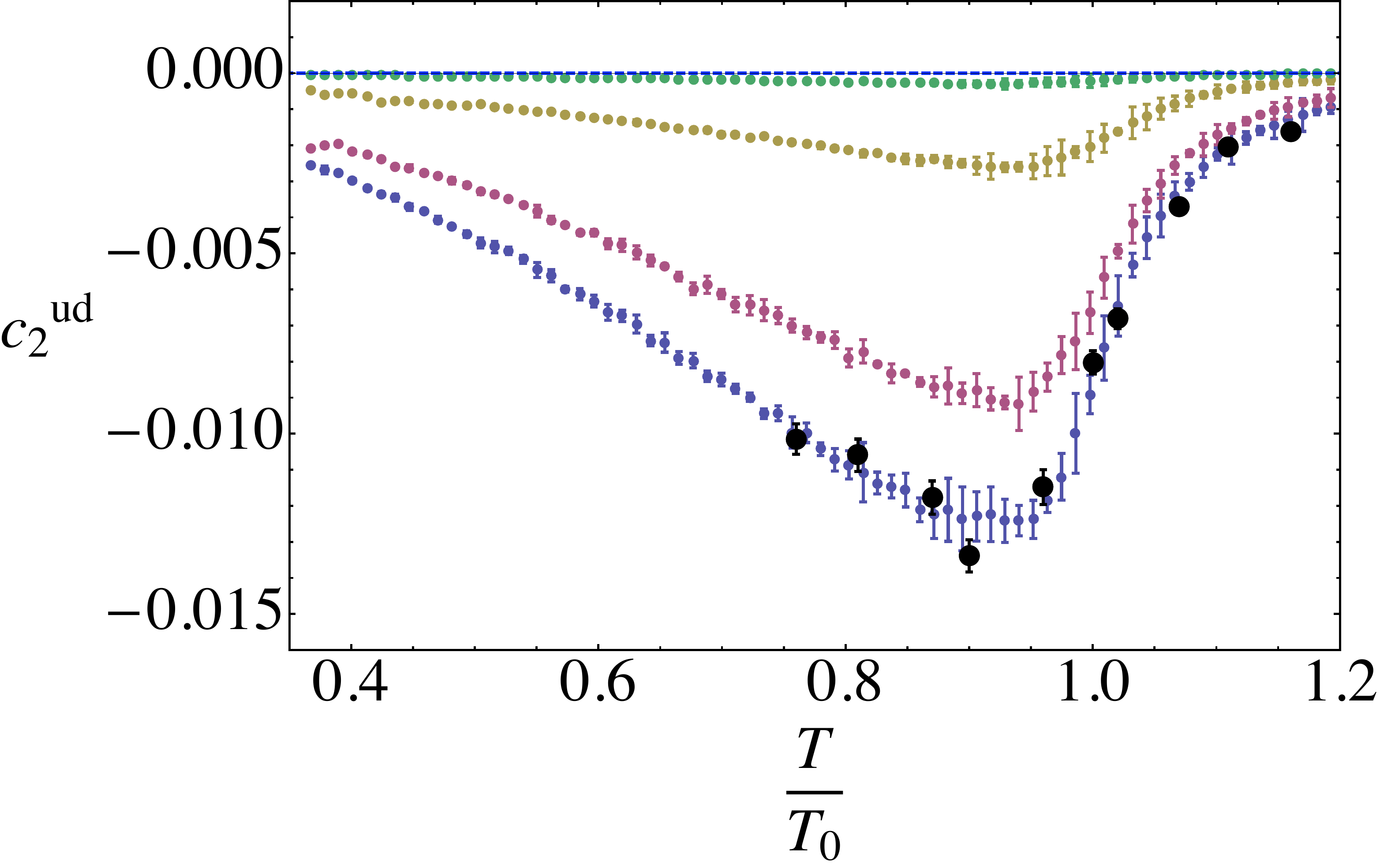}}
      \caption{Second order diagonal and non-diagonal coefficients $c_2^{uu}$ and $c_2^{ud}$. Black points are lattice data from \cite{Allton:2005gk}; the colored lines show the model predictions ($k$ increase from blue to red)}\label{fig:c2uuc2ud}
\end{figure}

The definition of the flavor diagonal and non-diagonal coefficients we are interested in is given by
\be
	c_n^{uu}=\frac{c_n+c_n^I}{4},&& c_n^{ud}=\frac{c_n-c_n^I}{4}.
\ee
In Fig.~\ref{fig:c2uuc2ud} we show the results for $c_2^{uu}$ and $c_2^{ud}$. For the diagonal coefficient we can see that the different choices of $k$ affect the matching with the lattice data (black points) only slightly. The picture changes totally when looking at the non-diagonal coefficient: in this case, at increasing volume the signal reduces. This is exactly what we expect because we know that with growing $V$ the fluctuations responsible for the non vanishing expectation value disappear. Thus we recover the vanishing mean-field result (dashed light-blue line). 

It is quite surprising, though, that for the smallest values of $k$ we can reproduce the lattice data with reasonable accuracy. This seems to indicate that the PNJL model can capture indeed the dynamics of the fluctuations at the origin of the non-vanishing $c_2^{ud}$.

\section{Conclusions}\label{sec:concl}
In this work we applied standard Monte-Carlo techniques in order to go beyond mean-field in the Polyakov loop extended Nambu -- Jona-Lasinio model calculations. Fluctuations introduced in this way strongly depend on a new parameter that fixes the minimal volume $\tilde{V}$ as defined in Sec.~\ref{sec:PNJLpf}. Inside this volume the fields are totally correlated while they are totally de-correlated on distances larger than $\tilde{V}^{1/3}$. Studying the QCD thermodynamics we found that the introduction of fluctuations does not alter the behavior of the traced Polyakov loop in the pure gauge sector nor the behavior of the Polyakov loop and chiral condensate in $N_f=2$. On the other hand these fluctuations are crucial in the evaluation of the Taylor expansion coefficients of the pressure. In particular, using the MC-PNJL we can reproduce very well the lattice data for the second non-diagonal coefficient which vanishes in mean field approximation.    

Collecting all this results we have that, the MC-PNJL model on the one hand includes correctly the dynamics associated with confinement and chiral symmetry breaking, on the other hand we can reproduce quite well lattice data. So it runs for as a good model in the investigation of the full phase diagram.
\\

\bibliography{PNJL}
\end{document}